\documentclass[11pt]{article}
\pdfoutput=1
\usepackage[preprint]{acl}
\usepackage{algorithm,algorithmicx,algpseudocode}
\usepackage{times}
\usepackage{latexsym}
\usepackage{listings}
\usepackage{iftex}
\usepackage[T1]{fontenc}
\usepackage[utf8]{inputenc}
\usepackage{times}

\usepackage{xcolor}
\usepackage{multirow}

\usepackage{microtype}

\usepackage{inconsolata}
\usepackage{amsmath, amssymb}
\usepackage{graphicx}

\title{Auto-Stega: An Agent-Driven System for Lifelong Strategy Evolution in LLM-Based Text Steganography}

\author{
  Jiuan Zhou\thanks{Equal contribution.} \\
  East China Normal University \\
  Shanghai, China
  \And
  Yu Cheng\footnotemark[1] \\
  East China Normal University \\
  Shanghai Innovation Institute \\
  Shanghai, China
  \AND
  Yuan Xie \\
  East China Normal University \\
  Shanghai Innovation Institute \\
  Shanghai, China
  \And
  Zhaoxia Yin\thanks{Corresponding author.} \\
  East China Normal University \\
  Shanghai, China
}

\begin{document}
\maketitle
\begin{abstract}

With the rapid progress of LLMs, high quality generative text has become widely available as a cover for text steganography. However, prevailing methods rely on hand-crafted or pre-specified strategies and struggle to balance efficiency, imperceptibility, and security, particularly at high embedding rates. Accordingly, we propose Auto-Stega, an agent-driven self-evolving framework that is the first to realize self-evolving steganographic strategies by automatically discovering, composing, and adapting strategies at inference time; the framework operates as a closed loop of generating, evaluating, summarizing, and updating that continually curates a structured strategy library and adapts across corpora, styles, and task constraints. A decoding LLM recovers the information under the shared strategy. To handle high embedding rates, we introduce PC-DNTE, a plug-and-play algorithm that maintains alignment with the base model's conditional distribution at high embedding rates, preserving imperceptibility while enhancing security. Experimental results demonstrate that at higher embedding rates Auto-Stega achieves superior performance with gains of 42.2\% in perplexity and 1.6\% in anti-steganalysis performance over SOTA methods.

\end{abstract}

\section{Introduction}

Steganography is an information hiding technique that embeds secret information in cover media to enable covert communication \cite{liao2025framework,11071645,10919187}. The fundamental objectives of steganographic systems are characterized by three key metrics: the embedding rate, which quantifies the capacity of secret information that can be hidden; perceptual concealment, typically represented by the quality of the media after information embedding (e.g., text quality or degree of image modification); and anti-steganalysis performance \cite{10872893,11030273}, which measures the capability to resist machine detection. These correspond respectively to core requirements of efficiency, imperceptibility, and security.

\begin{figure}[t]
  \centering
  \includegraphics[width=1.0\linewidth]{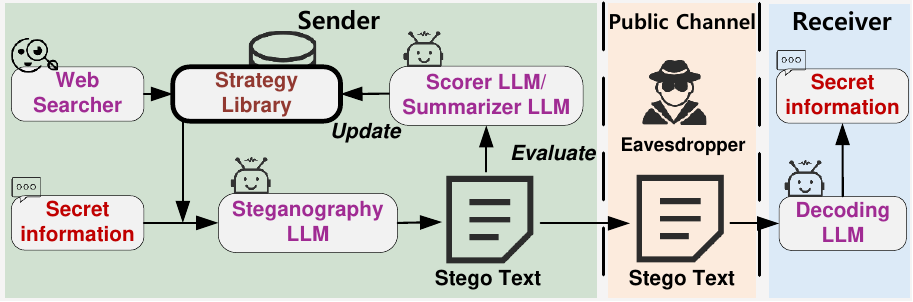} 
  \caption{The pipeline of Auto-Stega.}
  \label{fig:figure1}
\end{figure}

By cover type, steganography can be categorized into text steganography \cite{10904286,wang2025stlc,wu2024generative}, image steganography \cite{11005468,cheng2025rfnns}, audio steganography \cite{11036088,zhang2025triple}, and video steganography \cite{11071645}. Among these, text steganography has emerged as an active research area in information security, driven by the prevalence of text-based social communication.

Early text steganography was predominantly based on manually designed content modifications, such as synonym substitution \cite{li2019generating}, and spelling conversion \cite{shirali2008text}. 

Although these rule-based schemes can remain semantically unobtrusive, they usually provide a low embedding rate and induce a distributional change from natural text \cite{zhang-etal-2021-provably}, which compromises their resistance to machine detection and leads to inadequate anti-steganalysis performance.

\begin{figure*}[htbp]
  \centering
  \includegraphics[width=0.7\linewidth]{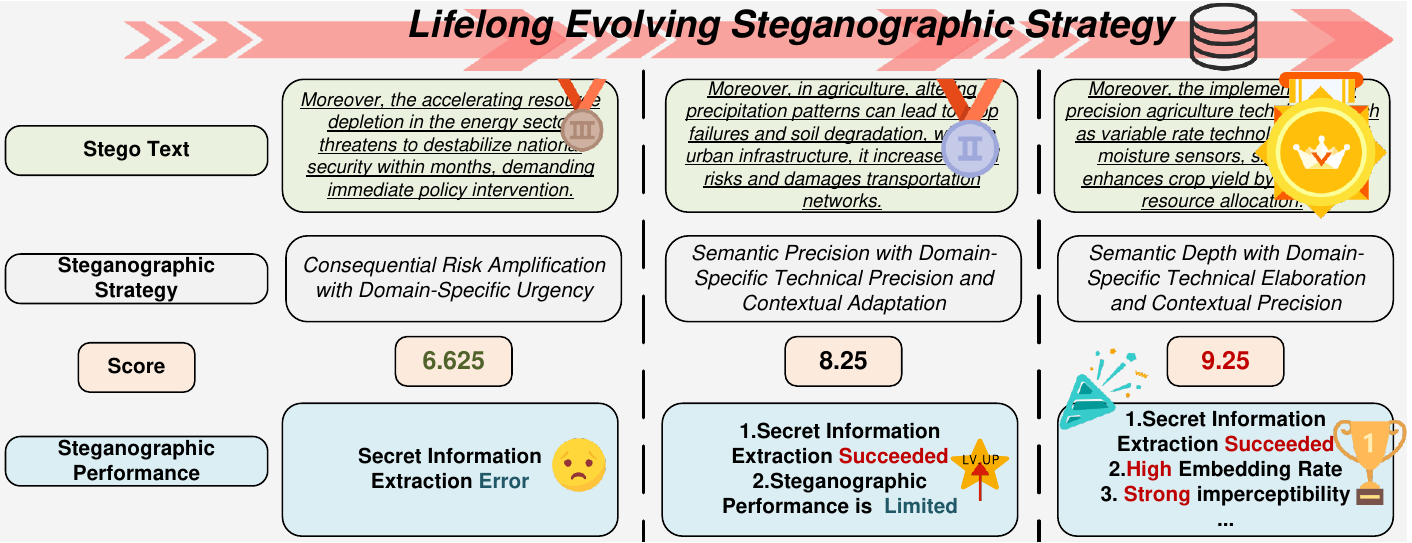} 
  \caption{Lifelong evolution of steganographic strategies.}
  \label{fig:figure2}
\end{figure*}

Recent advances in large language models (LLMs) have made high-quality generative text broadly accessible and controllable, creating new opportunities for generative text steganography. Generative steganography \cite{11005468,peng2024ldstega} is a technique that embeds information by guiding probabilistic generative models during content synthesis. In text generation, it typically embeds bits by sampling steering tokens within the model’s conditional probability distribution. Zhang et al. \cite{zhang-etal-2021-provably} achieve information-theoretic alignment via equal mass dynamic binning but rely on numerically fragile reconstruction; Ding et al. \cite{ding2023discop} strengthen security with distribution copies and seed-based decoding at the cost of synchronization and latency; Wu et al. \cite{wu2024generative} reduce shift using SVO and sentiment prompting, yet remain template and keyword dependent under domain shift. Taken together, these methods still rely on hand-crafted or pre-defined strategies and struggle to balance efficiency, imperceptibility, and security, especially under high embedding rates - a well-recognized challenge in the the steganography domain, known as the “trilemma.”

To fundamentally address this trilemma and inspired by cognitive science showing that creative solutions often emerge from coherently combining seemingly unrelated knowledge \cite{koestler1964act,benedek2012associative,lee2024empirical}, we propose Auto-Stega, an agent-driven, self-evolving framework (as shown in Fig. \ref{fig:figure1}) that is the first to realize self-evolving steganographic strategies by automatically discovering, composing, and adapting strategies at inference time. The framework unifies strategy retrieval, stego text generation, calibrated multi-objective evaluation, and strategy summarization with library update in a closed-loop, continually maintaining a searchable strategy library and coordinating efficiency, imperceptibility, and security across tasks. For high embedding rates, we also introduce PC-DNTE, an optional algorithm of our own design that integrates without changing the overall operation. The evolution of steganographic strategies is shown in Fig.~\ref{fig:figure2}; as the library is progressively updated, the generated stego text exhibits improved empirical performance. The contributions are as follows:

\begin{itemize}
\item \textbf{The first text steganography framework for self-evolving steganographic strategies.} It performs automatic strategy discovery and adaptation, coordinating efficiency, imperceptibility, and security through a closed loop comprising strategy retrieval, stego text generation, evaluation, and library updating.

\item \textbf{Lifelong evolving steganographic strategy.} A summarizer LLM codifies successful strategies into structured, searchable entries; retrieval based on score changes and subsequent reranking drive continual improvement and reliable transfer across corpora and task settings.
\item \textbf{Plug-and-play high embedding rates mapping.} We design PC-DNTE, an algorithm that preserves the base model’s conditional distribution while enabling distribution-preserving sampling at high embedding rates.

\item \textbf{Low-cost, requirement-oriented inference.} The pipeline runs entirely at inference time, without the need for training, fine-tuning or parameter access. It adapts to user defined requirements, delivering superior performance across various contexts.
\end{itemize}

\section{Related Work}

\subsection{Large Language Models}

LLMs are advanced neural networks trained on extensive text corpora, enabling them to generate and comprehend natural  language. These models have been employed as task-specific data generators for tabular data, relational triples and instruction data, achieving reasonable zero-shot quality under simple class-conditional prompts \cite{chia-etal-2022-relationprompt, schick-schutze-2021-generating}. 

Beyond data generation, LLMs are increasingly framed as engines for scientific discovery \cite{karpatne2025ai, luo2025llm4sr}. Tool-augmented, agentic systems illustrate this shift: ChemCrow \cite{m2024augmenting} links LLMs with expert chemistry tools, and Co-scientist \cite{boiko2023autonomous} automates experiment planning and cloud lab execution. Recent work has also explored hypothesis discovery and evaluation through systems like ResearchAgent \cite{baek-etal-2025-researchagent} and LiveIdeaBench \cite{ruan2024liveideabench}, which propose and refine ideas based on literature and single keyword prompts. Complementing these trends, jailbreak studies \cite{liudanturbo2025, zhou2025autoredteamer} demonstrate self-evolving jailbreak strategies, where agents autonomously discover and refine attacks over time.

\subsection{Generative Text Steganography}

Generative text steganography designs steganographic strategies that steer LLMs to select tokens so as to embed secret information within natural language. In this paradigm, a steganographic mapping associates bit strings with sampling decisions inside the model conditional token distribution, thereby controlling token selection by the LLM. From a provable security perspective, Zhang et al. \cite{zhang-etal-2021-provably} propose Adaptive Dynamic Grouping (ADG), which recursively groups vocabulary tokens to embed information while aligning with the base distribution under stated assumptions. Ding et al. \cite{ding2023discop} introduce "Distribution Copies" (Discop), a mapping designed to preserve the model’s original distribution during embedding. Wu et al. \cite{wu2024generative} implement a prompt based mapping that leverages interactions with LLM APIs to realize practical steganographic encoding. Despite these advances, current methods still face challenges in maintaining imperceptibility across high embedding rates and domains.

\section{Method}
\begin{figure*}[htbp]
  \centering
  \includegraphics[width=1.0\linewidth]{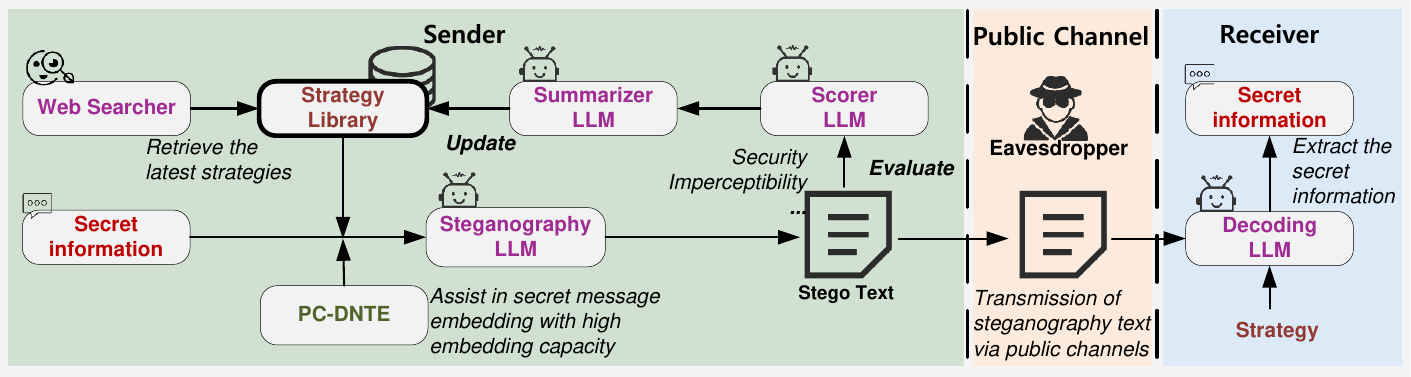} 
  \caption{The overall framework of Auto-Stega.}
  \label{fig:figure3}
\end{figure*}

Auto-Stega is an LLM-based text steganography framework that employs an agent-driven, self-evolving mechanism to continually update a strategy library and provide steganographic strategies tailored to the task. As shown in Fig. \ref{fig:figure3}, the system consists of a sender and a receiver. On the sender side, a closed loop of "generation, evaluation, summarization, and update" operates. The web searcher retrieves up-to-date strategies according to task requirements and updates the library; the steganography LLM then generates stego text, and the evaluation module, via the scorer LLM, assigns scores on multiple metrics, including efficiency, imperceptibility, and security. The summarizer LLM writes structured strategy entries back to the library, thereby realizing continuous strategy evolution. On the receiver side, the decoding LLM accurately recovers the secret information according to the shared steganographic strategy.

\begin{table}[t]
\centering
\small
\setlength{\tabcolsep}{6pt}        
\renewcommand{\arraystretch}{1.05} 
\newcommand{\ccell}[1]{\parbox[t]{\linewidth}{\centering #1}} 

\begin{tabular}{p{0.26\linewidth} p{0.68\linewidth}}
\hline
\ccell{\textbf{Notation}} & \ccell{\textbf{Description}} \\
\hline
\ccell{$T_c$} & \ccell{Cover text} \\
\ccell{$T_s$} & \ccell{Stego text} \\
\ccell{$B$} & \ccell{Number of bins} \\
\ccell{$S$} & \ccell{Score (scalar)} \\
\ccell{$\theta$} & \ccell{Fixed params of base AR LM} \\
\ccell{$m$} & \ccell{Target bin index} \\
\ccell{$S_T$} & \ccell{Score threshold} \\
\ccell{$R$} & \ccell{Evaluation response} \\
\hline
\end{tabular}
\caption{Notations}
\label{tab:notations}
\end{table}

\subsection{Web Searcher}

Starting from a library $L$ of text steganography strategies, the web searcher uses extensible literature retrieval and information extraction templates to mine recent work and convert candidate methods into executable entries. It evaluates each method with standardized metrics, assigns recommended scenarios, and compares it with existing entries; candidates that exceed the threshold $S_T$ are admitted to the lifelong evolving steganographic strategy library.

\subsection{Steganographic Generator and Evaluator}
Given a request $M$, the system first filters candidate strategies using the “Applicable Scenarios” field recorded in each library entry. It then performs multi-objective ranking of candidates based on evaluation metrics, selects the top-$k$ strategies, and uses steganography LLM to generate stego text. The resulting outputs are assessed by the evaluation module; if the predefined acceptance criteria are not met, generation is re-initiated with the next strategy in the ranked set, and the process iterates until the stopping condition is satisfied.

The evaluation module, comprising the detection LLM, which estimates the probability that a given text is steganographic and provides explanatory rationales, and other components, applies predefined criteria, including embedding rate, perplexity (PPL) \cite{mikolov2010recurrent}, and semantic similarity (SS). The embedding rate (ER) can be computed as:
\begin{equation}
\label{eq:EC}
ER = \frac{N_b}{W}\,
\end{equation}
where $N_b$ is the total number of bits in the embedded secret information, and $W$ is the total number of words in the stego text. PPL is a common quantitative measure in text generation, defined as:
\begin{equation}
\label{eq:ppl}
\mathrm{PPL} = \exp\!\left( -\frac{1}{N_w}\sum_{i=1}^{N_w} \log p\!\left(w_i \mid w_1,\ldots,w_{i-1}\right) \right).
\end{equation} where $N_w$ is the length of the text, $w_i$
is the $i$-th token in text, and $p(w_i\!\mid\!w_1,\ldots,w_{i-1})$ is the probability assigned by the language
model to the $i$-th word given the preceding words. The scorer LLM then compares the response $R$ and the score $S$ produced by the evaluation module against predefined criteria to determine whether the steganographic requirements are satisfied.

\begin{figure*}[htbp]
  \centering
  \includegraphics[width=0.9\linewidth]{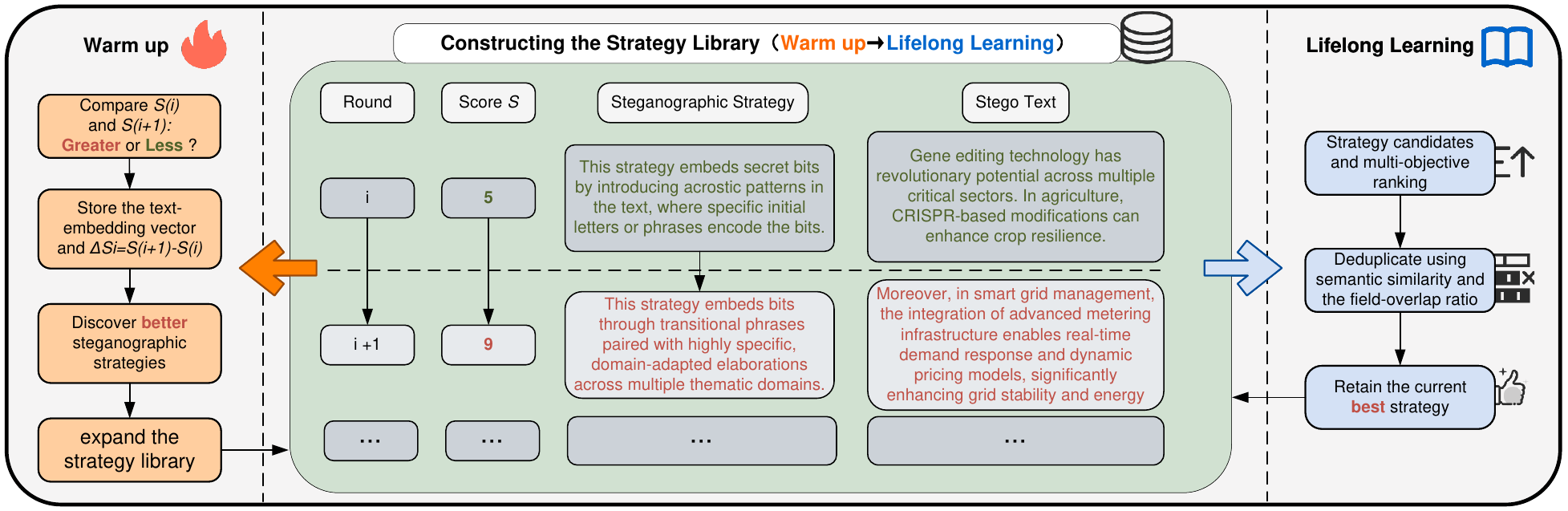} 
  \caption{Steganographic strategy library construction: from warm-up to lifelong learning.}
  \label{fig:figure4}
\end{figure*}

\subsection{Strategy Library Construction} 
The steganographic strategy library $L$ is organized into two stages: warm-up and lifelong learning. In the warm-up stage, we execute a generate–evaluate loop and log $(T_s, R, S)$ for each request. A summarizer LLM abstracts improvements between nearby or sampled records into structured strategy entries and writes them back to the library $L$ after deduplication. Each entry stores motivation, implementation notes, and evaluation summaries. The library $L$ is maintained as a key–value store whose keys $\kappa$ are response embeddings for efficient retrieval. Admission is gated by thresholded scores $S_T$, and operational details are given in Fig.~\ref{fig:figure4}. In the lifelong learning stage, given a request $M$ and current $L$, the system retrieves and ranks candidate strategies, regenerates by evolving new strategies from existing entries in $L$ when acceptance criteria are not met, admits improved entries under the same threshold rule, and terminates when scores exceed preset thresholds or the iteration budget is reached. Under the same thresholding rule, the system applies the single-best strategy when it clearly dominates the candidates. When several near-matching strategies jointly satisfy the acceptance criteria, the system assembles them to drive further evolution. When no candidate reaches $S_T$, the system discards the shortlist and initiates alternative discovery. The outputs are the ranked shortlist $\Gamma$ that is the candidate list constructed during strategy retrieval, and the selected set \textsc{EFFECTIVE}. The procedure is illustrated in Fig.~\ref{fig:figure4}. This compact process turns observed gains into reusable fragments and sustains continual evolution.

\algrenewcommand\algorithmiccomment[1]{\hfill{\footnotesize\textit{//~#1}}}
\algrenewcommand\algorithmicindent{0.8em} 

\algnewcommand\algorithmicabbrev{\textbf{Functions:}}
\algnewcommand\Functions{\item[\algorithmicabbrev]}

\begin{algorithm}[t]
\caption{Strategy Retrieval and Shortlisting}
\label{alg:retrieval-selection}
\scriptsize
\begin{algorithmic}[1]
  \Require Request $M$; library $L$ (entry $j$ with key $\kappa_j$); shortlist size $k$
  \Ensure Shortlist $\Gamma$
  \Functions \textproc{GE}: generate and evaluate (outputs $T_S$, $R$, $S$); \textproc{EMB}: embed $R$; \textproc{STS}: stable similarity top-$2k$; \textproc{MD}: metric discrepancy vs.\ current metrics; \textproc{TOPK}: pick $k$ by criterion

  \State $(T_S, R, S) \gets \textproc{GE}(M)$ \Comment{gen/eval}
  \State $E_R \gets \textproc{EMB}(R)$ \Comment{embed}
  \State $\mathcal{C} \gets \textproc{STS}(L, E_R, 2k)$ \Comment{retrieve}
  \ForAll{$j \in \mathcal{C}$}
    \State $d_j \gets \textproc{MD}(j, S)$ \Comment{score gap}
  \EndFor
  \State $\Gamma \gets \textproc{TOPK}(\mathcal{C}, \{d_j\}, k)$ \Comment{shortlist}
  \State \Return $\Gamma$
\end{algorithmic}
\end{algorithm}

\subsection{Strategy Retrieval}
Given a request $M$, the steganographic LLM and the evaluator module produce a stego text, a response, and an overall score $S$. We embed the response $R$ and retrieve the top-$2k$ entries from $L$ by similarity to their keys $\kappa_j$, with stable deterministic tie resolution. The candidates are reranked by metric discrepancy with the current metrics; the best $k$ form the shortlist $\Gamma$. Note that in the first iteration, no response is available for retrieval reference. Thus the steganography LLM is prompted without employing a steganographic strategy in the first iteration. Algorithm \ref{alg:retrieval-selection} formalizes the retrieval and selection procedure.

\subsection{Plug-and-Play High Embedding Rates Mapping}

Prior heuristics such as top-$p$ (nucleus) and locally typical sampling shape the candidate set near the model conditional distribution, improving fluency. However, they lack control over token selection that is calibrated for capacity and aware of the underlying distribution: per-step capacity remains uncalibrated, probability mass is unstructured, and the mapping from bits to tokens is unstable as candidate diversity varies. Accordingly, we propose PC-DNTE (Parity-Constrained Dynamic Nucleus-Typical Encoder).

\textsc{PC-DNTE} is a training free encoder. At each step, it adapts $p \in [p_{\min}, p_{\max}]$ to form a nucleus candidate set under $p_{\theta}(\cdot \mid \mathrm{ctx}_t)$ and applies a local typicality filter; the filtered set is then partitioned into $B=32$ equal mass bins $\{G_m\}_{m=0}^{31}$. The choice of $B=32$ ensures that each bin represents a manageable fraction of the total probability mass, allowing the use of a 5 bit index for efficient token selection while preserving the base distribution. The encoder reads $5$ ciphertext bits to choose the bin index $m$ and one parity bit $b^{*}$, and selects tokens in $G_m$ whose seed conditioned parity matches $b^{*}$, yielding deterministic, seed synchronized decoding. If no such token exists, the algorithm expands $p$ (within $[p_{\min}, p_{\max}]$) and retries; if still empty, it records a no embed step, defers the bits, and emits $\arg\max_{y\in G_m} p_{\theta}(y \mid \mathrm{ctx}_t)$. The context is updated after emission. Near sentence boundaries, a small log odds boost to the end of sequence token (EOS) and punctuation together with a bounded closure routine encourages well formed breaks. This adjustment is applied after token emission and also during any no embed closure, and it only reweights the next step candidate set; consequently, it does not modify the bits already embedded. Taken together, the procedure preserves the model conditional distribution, provides deterministic synchronization via seed conditioned parity, and sustains high embedding rates through equal mass binning, with step by step operations in Algorithm \ref{alg:pc-dnte}.

\subsection{Secret Information Extraction} 
In the secret extraction module, when configured with the same steganographic strategy as the encoding side, the decoding LLM is able to accurately decode the embedded secret information.

\algrenewcommand\algorithmiccomment[1]{\hfill{\footnotesize\textit{//~#1}}}
\algrenewcommand\algorithmicindent{0.8em} 
\algnewcommand\algorithmicfunctions{\textbf{Functions:}}

\begin{algorithm}[t]
\caption{PC-DNTE}
\label{alg:pc-dnte}
\scriptsize
\begin{algorithmic}[1]
\Require conditional $p_{\theta}(\cdot \mid \mathrm{ctx})$; ciphertext \textsc{Bits}; $p_{\min},p_{\max}\!\in\!(0,1]$; bins $B{=}32$; scorer \textproc{TYP}$(\cdot)$; seed \textsc{Seed}; closure $\varepsilon$; budget $K$; punctuation set $\mathcal{P}$
\Ensure stego text $T_S=(t_1,\ldots,t_T)$
\Functions \textproc{AdaP}: adapt $p\!\in\![p_{\min},p_{\max}]$; \textproc{DynNuc}: nucleus under $p_{\theta}$ at $p$; \textproc{LocTyp}: local typicality filter; \textproc{EqM}: split to $B$ equal-mass bins; \textproc{ReadBits}: pop $n$ bits; \textproc{PH}: seed-conditioned parity; \textproc{IncP}: enlarge $p$ up to $p_{\max}$; \textproc{NoEmb}: record no-embed; \textproc{QueueBits}: enqueue bits; \textproc{UpdCtx}: update context; \textproc{SBLikely}: boundary test; \textproc{BoostEOS}: boost $\mathcal{P}$; \textproc{CloseSteps}: bounded closure

\State $T_S\gets[\,]$;\ $t\gets1$
\While{\textproc{NotEOM}(\textsc{Bits})}
  \State $\textsc{no\_emb}\gets\textsc{false}$
  \State $p\gets\textproc{AdaP}(p_{\min},p_{\max},\,p_{\theta}(\cdot\,|\,\mathrm{ctx}_t))$ \Comment{adapt}
  \State $C_t\gets\textproc{DynNuc}(p_{\theta}(\cdot\,|\,\mathrm{ctx}_t),p)$;\ $C'_t\gets\textproc{LocTyp}(C_t,\mathrm{ctx}_t,\textproc{TYP})$;\ $\{G_0,\ldots,G_{31}\}\gets\textproc{EqM}(C'_t,B)$
  \State $m\gets\textproc{ReadBits}(\textsc{Bits},5)$;\ $b^\ast\gets\textproc{ReadBits}(\textsc{Bits},1)$
  \State $S\gets\{\,y\in G_m:\ \textproc{PH}(y,\mathrm{ctx}_t,\textsc{Seed})=b^\ast\,\}$

  \If{$S=\varnothing$}
    \State $p'\gets\textproc{IncP}(p,p_{\max})$
    \State $C_t\gets\textproc{DynNuc}(p_{\theta}(\cdot\,|\,\mathrm{ctx}_t),p')$;\ $C'_t\gets\textproc{LocTyp}(C_t,\mathrm{ctx}_t,\textproc{TYP})$;\ $\{G_0,\ldots,G_{31}\}\gets\textproc{EqM}(C'_t,B)$
    \State $S\gets\{\,y\in G_m:\ \textproc{PH}(y,\mathrm{ctx}_t,\textsc{Seed})=b^\ast\,\}$
  \EndIf

  \If{$S=\varnothing$}
    \State \textproc{NoEmb}();\ \textproc{DefBits}($\{m,b^\ast\}$) \Comment{enqueue bits}
    \State $y_t\gets\arg\max_{y\in G_m}p_{\theta}(y\,|\,\mathrm{ctx}_t)$; 
    \State $\textsc{no\_emb}\gets\textsc{true}$ \Comment{fallback}
  \Else
    \State $y_t\gets\arg\max_{y\in S}p_{\theta}(y\,|\,\mathrm{ctx}_t)$ \Comment{greedy}
  \EndIf

  \State $T_S\gets T_S\cup\{y_t\}$;\ $\mathrm{ctx}_{t+1}\gets\textproc{UpdCtx}(\mathrm{ctx}_t,y_t)$;\ $t\gets t+1$
  \If{$\textproc{SBLikely}(\mathrm{ctx}_t,\varepsilon)\ \lor\ \textsc{no\_emb}$}
    \State \textproc{BoostEOS}($\mathcal{P},\varepsilon$);\ \textproc{CloseSteps}($\mathrm{ctx}_t,K$) \Comment{reweight}
  \EndIf
\EndWhile
\State \Return $T_S$
\end{algorithmic}
\end{algorithm}

\section{Experiment}
In this section, we evaluate the performance of Auto-Stega in terms of efficiency, imperceptibility, and security and human evaluation. The details of the experimental setup and result analysis are presented in the following subsections. Appendix \ref{appendix2} provides representative high scoring strategy entries, and Appendix \ref{appendix3} presents complete prompt templates for the steganographic LLM and the summarizer LLM.

\begin{table*}[t]
  \centering

  \setlength{\tabcolsep}{7pt}
  \renewcommand{\arraystretch}{1.10}
  \small

  \newcommand{\thickhline}{\noalign{\hrule height 1.2pt}}
  \newcommand{\midline}{\noalign{\hrule height 0.5pt}}

  \begin{tabular}{c ccc ccc ccc}
    \thickhline
    \multirow{2}{*}{\textbf{Methods}}
      & \multicolumn{3}{c}{News}
      & \multicolumn{3}{c}{Movie}
      & \multicolumn{3}{c}{Sentiment} \\
    \cline{2-10}
      & ER\(\uparrow\) & $\mathrm{PPL}^{\ast}$\(\downarrow\) & SS\(\uparrow\) 
      & ER\(\uparrow\) & $\mathrm{PPL}^{\ast}$\(\downarrow\) & SS\(\uparrow\) 
      & ER\(\uparrow\) & $\mathrm{PPL}^{\ast}$\(\downarrow\) & SS\(\uparrow\)  \\
    \midline
    ADG         & 4.99 & 8.58 & 0.40 & 5.44 & 5.39 & 0.45 & 5.22 & {\color{red}\textbf{0.53}} & 0.39 \\
    Discop     &  4.62  & 0.70 & 0.58 & 4.94 & 0.62 & 0.58 & 5.41 & 0.94  & 0.48 \\
    LLM-Stega  &  5.94 & 0.46 & 0.61 & 4.63  & 0.28 & 0.61 & 7.90 & 0.80 & 0.58 \\
    Ours       & {\color{red}\textbf{5.97}} & {\color{red}\textbf{0.01}} & {\color{red}\textbf{0.63}}
               & {\color{red}\textbf{5.69}} & {\color{red}\textbf{0.04}} & {\color{red}\textbf{0.62}}
               & {\color{red}\textbf{8.14}} & 0.84 & {\color{red}\textbf{0.60}} \\
    \thickhline
  \end{tabular}
\caption{Experimental results of $\mathrm{PPL}^{\ast}$ and SS under high embedding rates. ↑ higher is better, ↓ lower is better.}
\label{tab:main_results}
\end{table*}

\subsection{Experimental Setup}
{\bfseries Datasets.}
We evaluated Auto-Stega in three public corpora: the News Category Dataset (News) \cite{misra2022news}, the Large Movie Review dataset (Movie) \cite{maas2011learning}, and Sentiment140 (Tweet) \cite{go2009twitter}. For quantitative comparability, we use 12{,}300 News items (300 per category across 41 categories), 20{,}000 Large Movie Review entries, and 20{,}000 Sentiment140 tweets.
The combination spans domains, styles, and lengths, which allows us to assess efficiency, imperceptibility, and security under cross-scenario distribution shifts while maintaining sufficient scale for reliable estimation.

{\bfseries Baselines.}
We compare Auto-Stega with three state-of-the-art (SOTA) baselines, including ADG \cite{zhang-etal-2021-provably}, Discop \cite{ding2023discop}, and LLM-Stega \cite{wu2024generative}. To demonstrate that our method maintains strong performance across embedding capacities, we conduct experiments under both high embedding rates and low embedding rates settings and compare the stego text generated by different methods in each case.

{\bfseries Hyperparameters.} Evaluation was conducted in two stages. An initial warm up exploration was performed on 50 steganographic requests, running $N=150$ iterations to build the initial strategy library. Using this library, the runtime lifelong learning phase was then executed: for each request in the experimental datasets, 5 iterations were carried out. A full steganography round is defined as generating stego text for a given cover text and completing the evaluation. Unless otherwise stated, the maximum iteration budget was set at $T=150$ and the stopping threshold to $S_T=8.5$ for each data instance. During evaluation, the strategy library was frozen and an additional stego generation pass was conducted on the three datasets. For plug-and-play mapping, \( p_{\mathrm{min}} = 0.88 \) and \( p_{\mathrm{max}} = 0.95 \) were used to obtain sufficient candidate tokens while suppressing low typicality tails, and GPT-2 was used to compute next-token probabilities. We use \textsc{DeepSeek}\mbox{-}\textsc{V3.2} for the steganography, decoding, and scorer LLMs, and \textsc{GPT}\mbox{-}4\textsc{o} for the summarizer LLM.

\begin{table}[t]
  \centering

  \begingroup
  \setlength{\tabcolsep}{2.2pt}   
  \renewcommand{\arraystretch}{1.10}
  \footnotesize                  

  \newcommand{\thickhline}{\noalign{\hrule height 1.2pt}}
  \newcommand{\midline}{\noalign{\hrule height 0.5pt}}

  \begin{tabular}{c cc cc cc}
    \thickhline
    \multirow{2}{*}{\textbf{Methods}}
      & \multicolumn{2}{c}{News}
      & \multicolumn{2}{c}{Movie}
      & \multicolumn{2}{c}{Sentiment} \\
    \cline{2-7}
      & $\mathrm{PPL}^{\ast}$\(\downarrow\) & SS$\uparrow$
      & $\mathrm{PPL}^{\ast}$\(\downarrow\) & SS$\uparrow$
      & $\mathrm{PPL}^{\ast}$\(\downarrow\) & SS$\uparrow$ \\
    \midline
    ADG
      & 9.26 & 0.45
      & 5.92  & 0.49
      & {\color{red}\textbf{0.39}} & 0.38 \\
    Discop
      & 0.90 & 0.60
      & 0.88 & 0.59
      & 0.99 & 0.44 \\
    LLM\text{-}Stega
      & 0.03 & 0.64
      & 0.11  & 0.67
      & 0.94  & 0.58 \\
    Ours
      & {\color{red}\textbf{0.01}} & {\color{red}\textbf{0.66}}
      & {\color{red}\textbf{0.08}} & {\color{red}\textbf{0.72}}
      & 0.86 & {\color{red}\textbf{0.65}} \\
    \thickhline
  \end{tabular}
  \endgroup
\caption{Experimental results of $|\Delta\text{PPL}|$ and SS at a low embedding rate of 0.1 bpw.}
\label{tab:main_results_compact}
\end{table}

{\bfseries Efficiency.}
In the field of steganography, the embedding rate serves as a key evaluation metric to assess the efficiency of a method.
In text steganography, the embedding rate is defined as the average number of secret bits embedded per word (bits per word, bpw), as shown in Formula \ref{eq:EC}.

{\bfseries Imperceptibility.}
In text steganography, imperceptibility is typically categorized into text quality and statistical imperceptibility. Text quality captures human-perceived naturalness, including fluency, coherence, and semantic fidelity relative to the cover text. Statistical imperceptibility assesses the distributional alignment with the cover. Together, these facets provide a measure of whether a method conceals information without degrading readability or altering the underlying statistical characteristics of the text.

{\bfseries (1) Text Quality.}
PPL and SS are commonly used to evaluate the text quality of generated stego text. In this experiment, we choose the GPT-2 model to calculate the PPL values of different stego text. Prior work \cite{yang2018rnn} indicates that social media text exhibits substantial stylistic variability, so human-written sentences deviate widely from any single language model; consequently, stego text whose PPL is closer to that of human text is considered more secure. 
To compare across corpora with different baseline perplexities, we use a normalized perplexity deviation, $\mathrm{PPL}^{\ast}=\frac{\lvert \mathrm{PPL}_{\text{stego}}-\mathrm{PPL}_{\text{cover}}\rvert}{\mathrm{PPL}_{\text{cover}}}$ where lower values indicate better text quality. For the SS, we choose the Sentence-bert \cite{reimers-gurevych-2019-sentence} method and use the
roberta-base-nli-mean-tokens \cite{liu2019roberta} model to extract sentence vectors, to calculate the cosine similarity between the stego text and the cover text.

{\bfseries (2) Statistical Imperceptibility.}
Kullback-Leibler Divergence (KLD) serves as an evaluation metric to measure the statistical imperceptibility of steganographic algorithms by comparing the distribution of the generated stego text with the distribution of the cover text. In our experiments, we adopt the KLD proposed
by Zhang et al. \cite{zhang-etal-2021-provably} to evaluate the statistical imperceptibility of the methods tested. It is formulated as follows:
\begin{equation}
\label{eq:kld}
\resizebox{.98\linewidth}{!}{$
\mathrm{KLD}(\mu_x,\sigma_x,\mu_y,\sigma_y)
= \sum \Bigl[
\log\!\Bigl(\frac{\sigma_y}{\sigma_x}\Bigr)
+ \frac{\sigma_x^{2}+(\mu_x-\mu_y)^{2}}{2\sigma_y^{2}}
- \frac{1}{2}
\Bigr]
$}
\end{equation}

Where $\mu_x$ and $\sigma_x$ are the mean and standard deviation of the cover text, while $\mu_y$ and $\sigma_y$ represent those of the stego text.

{\bfseries Security.} 
Steganography and steganalysis are locked in an ongoing adversarial game. As such, anti-steganalysis performance serves as the paramount metric for evaluating the security of a steganographic algorithm. In this experiment, we leverage three advanced steganalysis methods, including LS-CNN (LC) \cite{wen2019convolutional}, BiLSTM-Dense (BD) \cite{yang2020linguistic}, and Bert-FT (BF) \cite{peng2021real}. The results are reported in terms of steganalysis accuracy—defined as the rate of correctly identifying whether a given text contains hidden information—where values closer to  $50\%$ (equivalent to random guessing) indicate stronger anti-steganalysis performance.

{\bfseries Human Evaluation.}
For the human evaluation, we focus on evaluating three key aspects of the generated texts: fluency, coherence, and relevance. Our human evaluation was conducted by five researchers who were not involved in the study, and each one holds a master's degree. The evaluation process involved a mixed assessment sheet containing both generated stego and cover text; only stego text scores were selected for analysis. The researchers, unaware of the underlying algorithms, labels, or any other details, were tasked solely with assessing and scoring text quality. Each researcher independently rated the texts on a five-point scale (ranging from ’very poor’ to ’very good’). The higher evaluation score denotes the better generated stego text.

\begin{table}[t]
  \centering
  \begingroup
  \setlength{\tabcolsep}{7pt}
  \renewcommand{\arraystretch}{0.9}
  \newcommand{\thickhline}{\noalign{\hrule height 1.2pt}}
  \newcommand{\midline}{\noalign{\hrule height 0.5pt}}

  \resizebox{\linewidth}{!}{%
  \begin{tabular}{c c c c c}
    \thickhline
    \multirow{2}{*}{\textbf{Payload}} & \multirow{2}{*}{\textbf{Methods}}
      & \multicolumn{1}{c}{\textbf{News}}
      & \multicolumn{1}{c}{\textbf{Movie}}
      & \multicolumn{1}{c}{\textbf{Sentiment}} \\
    \cline{3-5}
      &  & \textbf{KLD}\(\downarrow\) & \textbf{KLD}\(\downarrow\) & \textbf{KLD}\(\downarrow\) \\
    \midline
    \multirow{4}{*}{Low}
      & ADG              & 2.63                     & 2.28                     & {\color{red}\textbf{1.90}} \\
      & Discop           & 2.78                     & {\color{red}\textbf{1.94}}        & 2.17 \\
      & LLM\text{-}Stega & {\color{red}2.41}                     & 2.83                     & 2.43 \\
      & Ours             & {\color{red}\textbf{2.21}} & {\color{red}2.13}       & {\color{red}2.11} \\
    \midline
    \multirow{4}{*}{High}
      & ADG              & 2.63                     & 2.27                     & {\color{red}\textbf{1.89}} \\
      & Discop           & {\color{red}1.94}        & {\color{red}\textbf{1.96}}        & 2.17 \\
      & LLM\text{-}Stega & 2.33                     & 2.22                     & 2.19 \\
      & Ours             & {\color{red}\textbf{1.87}} & {\color{red}1.98}       & {\color{red}2.05} \\
    \thickhline
  \end{tabular}%
  }
  \endgroup
 \caption{The experimental results of the statistical imperceptibility under high and low embedding rates. The best results are highlighted in bold red, and the second bests  are shown in red without bold. \(\downarrow\) lower is better.}
\label{tab:kld_payload_fit}
\end{table}

\subsection{Results and Analysis}
{\bfseries (1) Efficiency and Imperceptibility.}
In generative text steganography, efficiency and imperceptibility are intrinsically linked: changes to the embedding rate (ER) modify sampling behavior and can materially affect both text quality and statistical imperceptibility. Accordingly, we analyze efficiency and imperceptibility jointly in this section.

\textit{Text Quality}. The experimental results for text quality and ER are listed in Tables \ref{tab:main_results} and \ref{tab:main_results_compact}. Auto-Stega achieves a higher ER while delivering superior performance. In our experiments, the average PPL of cover sentences is 113.89 on the News corpus, 111.12 on the Movie corpus, and 1224.79 on the Tweet corpus. Auto-Stega achieves the highest SS across all datasets and attains the lowest $\mathrm{PPL}^{\ast}$ on News and Movie. On Tweet, the stego text exhibit a slightly higher $\mathrm{PPL}^{\ast}$; nevertheless, Auto-Stega achieves a 42.2\% relative reduction compared with the SOTA baseline. These corpus characteristics highlight the importance of maintaining stylistic consistency with the cover text, thereby improving overall textual quality.  Within Auto-Stega, PC-DNTE treats text quality as an explicit objective during generation and selection, thereby generating high quality stego text.

\textit{Statistical Imperceptibility.} In this part, we compute sentence-level KLD using distributions derived from a pre-trained BERT encoder, comparing the cover and stego sets. In our experiments, we take the logarithm of the KLD values to facilitate a clearer comparison of statistical imperceptibility between algorithms. As shown in Table \ref{tab:kld_payload_fit}, under both high and low embedding rates, Auto-Stega achieves the lowest KLD on the News dataset, and the second lowest KLD on the Movie and Tweet corpora. ADG \cite{zhang-etal-2021-provably} performs well on tweets, whereas Discop \cite{ding2023discop} performs well on reviews, because their equiprobable grouping and distribution preserving sampling better match the peaked token distributions and formulaic phrasing characteristic of these datasets. In addition, our method can be configured to incorporate sentiment conditioning or template-based generation, which can reduce KLD on these two corpora while preserving semantic fidelity.

Taken together, these results demonstrate that Auto-Stega achieves the strongest overall imperceptibility across embedding rates, indicating that it maintains superior perceptual concealment even under high embedding rates and is well-suited for practical deployment in real-world applications.

\begin{figure}[t]
  \centering
  \includegraphics[width=0.9\linewidth]{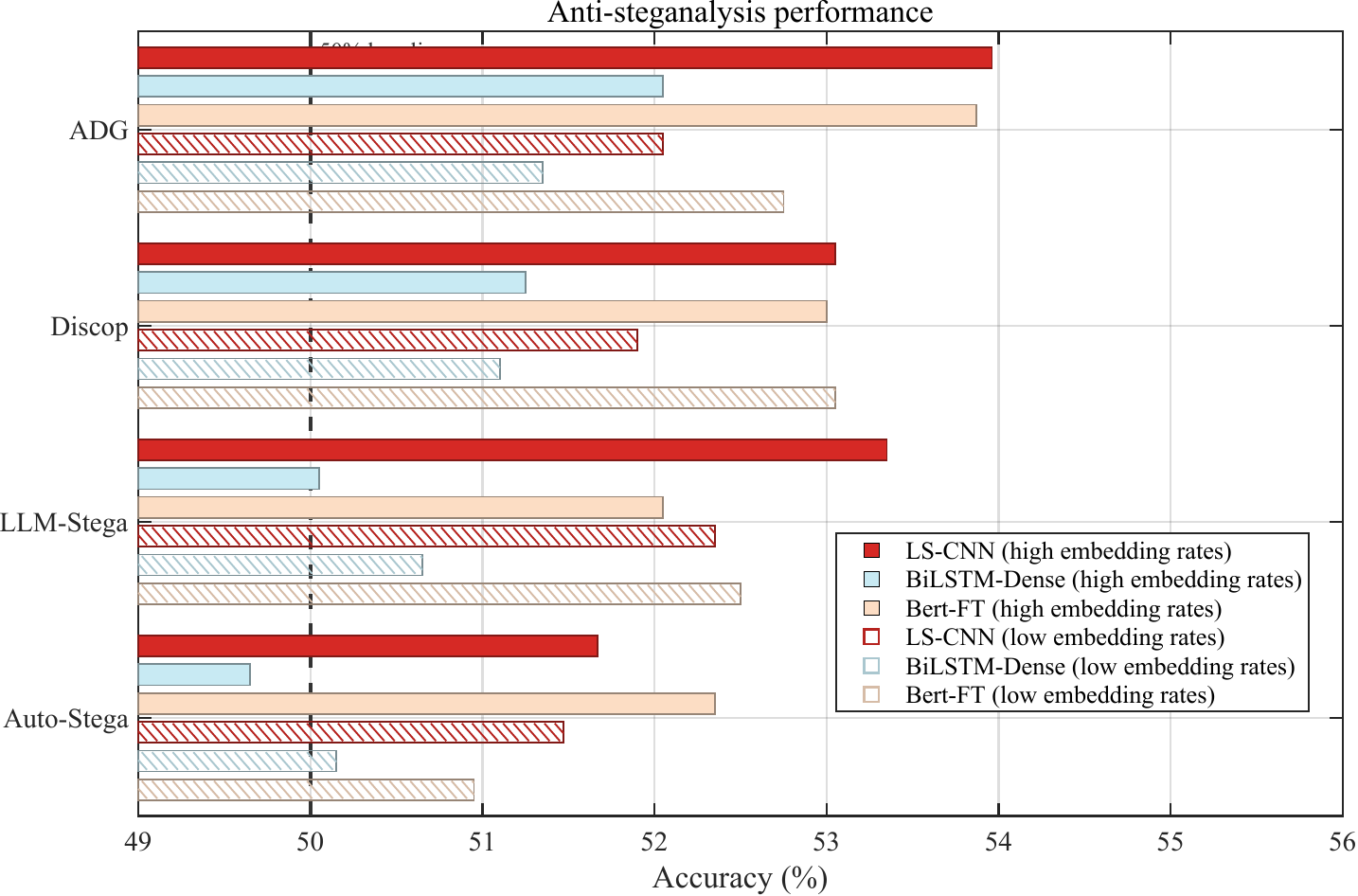}
  \caption{The results of the anti-steganalysis performance across low and high embedding rates.}
  \label{fig:anti}
\end{figure}

{\bfseries (2) Security.}
We used splits of 10{,}000 training pairs, 1{,}000 validation pairs, and 1{,}000 test pairs, each pair containing a cover text and its corresponding stego text. Fig. \ref{fig:anti} presents experimental results on anti-steganalysis performance across different embedding rates. The proposed method, Auto-stega, continues to exhibit superior performance across all metrics when compared to other steganography methods. Notably, Auto-Stega achieves LC of 51.67\%, BD of 49.65\%, and BF of 52.35\%, representing an average 1.6\% reduction relative to the SOTA method. This improvement reflects that, during iterative generation and selection, Auto-Stega explicitly evaluates anti-steganalysis performance and admits candidates that score better on this criterion, thereby producing stego text that is harder to detect. Overall, these results indicate that Auto-Stega achieves the strongest security among evaluated methods.

\begin{figure}[t]
  \centering
  \includegraphics[width=0.8\linewidth]{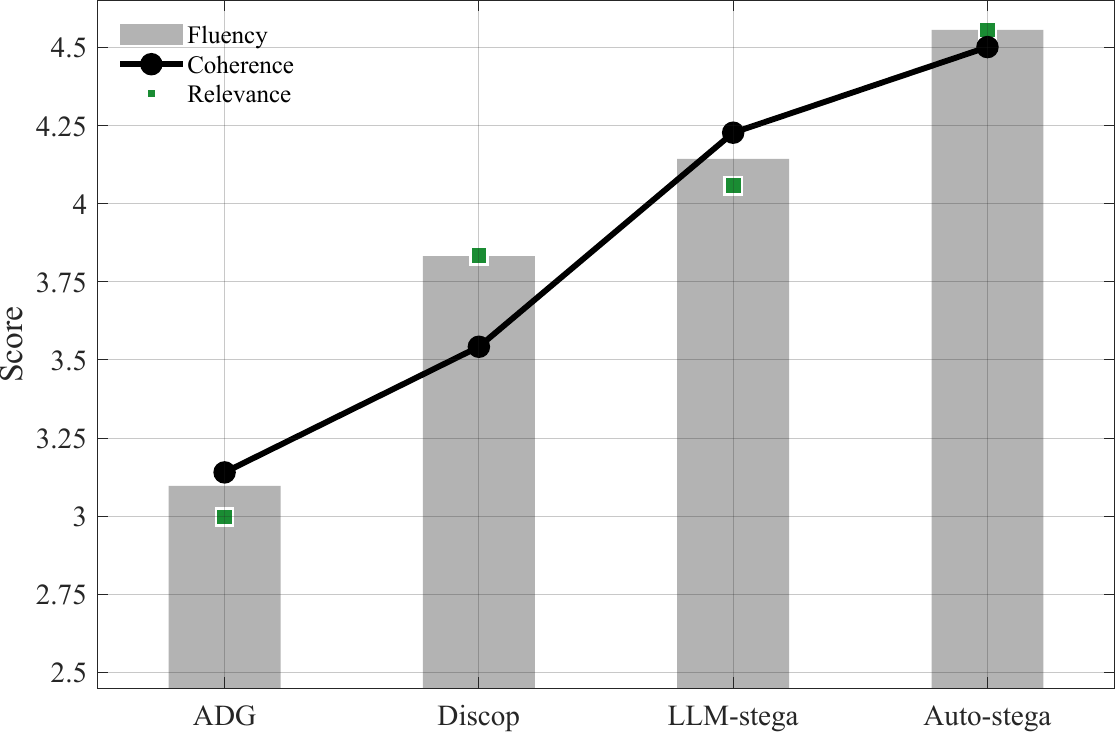}
  \caption{The results of the human evaluation.
}
  \label{fig:human}
\end{figure}

{\bfseries (3) Human Evaluation.}
To further assess the effectiveness of the proposed Auto-Stega, we conducted a human evaluation, with detailed results shown in Figure~\ref{fig:human}. Human ratings demonstrate that Auto-Stega outperforms the three baseline methods, producing stego text that is more fluent, coherent, relevant, and more imperceptible. This finding is particularly significant in the context of text steganography, underscoring the potential of Auto-Stega in applications where information concealment is critical. Moreover, as shown in Appendix~\ref{appendix1} Table~\ref{tab:stego_examples}, we randomly selected several examples of stego text generated by Auto-Stega and the baselines for qualitative analysis. The results indicate that Auto-Stega generates texts with high fluency, grammatical correctness, and semantic coherence.

\section{Conclusion}
We presented Auto-Stega, an agent-driven, self-evolving framework that discovers, composes, and adapts steganographic strategies entirely at inference time. By continually curating a searchable strategy library without retraining and by optionally employing the plug-and-play PC-DNTE to preserve the base model’s conditional distribution, the system supports higher embedding rates while maintaining imperceptibility and security. Experiments demonstrate a consistently superior trade-off among efficiency, imperceptibility, and security compared with SOTA baselines. Future work will extend the framework to multilingual and multimodal covers, strengthen robustness against adaptive steganalyzers and noisy channels, and reduce inference latency.

\section*{Limitations}
\label{sec:limitations}
While Auto-Stega achieves efficient covert text communication across diverse textual domains, it does not yet address multimodal settings in which information may be distributed across heterogeneous cover media. As future work, we will extend Auto-Stega to multimodal steganography with multiple covers by integrating modality aware generators and evaluators with the strategy library, enabling coordinated cover selection, adaptive embedding rate allocation, and cross modal redundancy. This extension is expected to substantially improve practical utility and covertness in real world deployments.

\bibliography{arxiv}

\clearpage
\appendix

\section{Appendix}
\label{sec:appendix}

\subsection{Comparison of Steganographic Examples Generated by Proposed Method and Existing Methods.}
In this section, we analyze randomly sampled stego text generated by Auto-Stega and three baselines (ADG, Discop, and LLM-Stega). As shown in Table \ref{tab:stego_examples}, texts from ADG and Discop frequently exhibit surface issues, such as awkward collocations, agreement errors, and missing function words, especially near high entropy positions induced by their mappings. LLM-Stega generally maintains sentence level fluency, but shows intersentential drift and occasional semantic inconsistencies at higher embedding rates, reflecting template and keyword rigidity. In contrast, Auto-Stega produces texts that are fluent, syntactically well formed, and semantically consistent across sentences, while more closely matching the style of the source corpus (for example, concise headlines in News and an informal register in Tweets). Manual inspection also finds fewer detectability cues, consistent with the smallest difference from cover perplexity observed in the experimental results. Overall, the results indicate that Auto-Stega yields more coherent and less detectable stego text than the compared methods.
\label{appendix1}

\begin{table*}[t]
  \centering
  \begingroup
  \setlength{\tabcolsep}{6pt}
  \renewcommand{\arraystretch}{1.95}

  \newcommand{\thickhline}{\noalign{\hrule height 1.2pt}}

  \newcommand{\stegoboxC}[2]{\parbox[c]{#1}{\centering #2}}   
  \newcommand{\stegoboxL}[2]{\parbox[c]{#1}{\raggedright #2}} 

  \begin{tabular}{cc}
    \thickhline
    \stegoboxC{0.22\linewidth}{\textbf{Methods}} &
    \stegoboxC{0.74\linewidth}{\textbf{Stego text example}} \\
    \hline

    \stegoboxC{0.22\linewidth}{ADG} &
    \stegoboxL{0.74\linewidth}{putting for me can eat me can go well} \\
    \hline

    \stegoboxC{0.22\linewidth}{Discop} &
    \stegoboxL{0.74\linewidth}{%
      I'm a Ser, its well over 20 years old and I just started to play. And thats it :D
      I'd be happy to give back to you please respond up till hour 30 More bugs I'm
      sure around here there are no edit black on black bars? See below. As if it
      wasn't clear..in game this spreadsheet just explains what blocks are. Although
      it didn't make clear to me what kind of bones were I carrying. When you cast
      it double kill yourself as if you. //
      ``Do you want to... trust me?...'' ``No... everyone IS right. I definitely agree,
      even just... wanting a girl you have never known.'' \{K...center, cums, gesture
      one more time\}......I trust your feelings, you really have.'' In other words
      it wasn't a grown up. Shame. DISCONTINUED //
      ``Ah, Engh...huh. Thank you, am I right...?'' ``Yes. Your Majesty, on behalf} \\
    \hline

    \stegoboxC{0.22\linewidth}{LLM-Stega} &
    \stegoboxL{0.74\linewidth}{Officials announce an exciting new team that promises to bring fresh
      talent and enthusiasm to the entertainment industry.} \\
    \hline

    \stegoboxC{0.22\linewidth}{Ours} &
    \stegoboxL{0.74\linewidth}{The future might be ready: robots like computers in some scenarios
      would rapidly replace humans.} \\
    \thickhline
  \end{tabular}
  \endgroup
\caption{Stego text examples produced by different methods.}
\label{tab:stego_examples}
\end{table*}

\subsection{Evolution of Steganographic Strategies}
In this section, we present steganographic strategy entries from the library spanning a range of scores. Each entry follows a standardized schema comprising \emph{Strategy}, \emph{Definition}, \emph{Technique}, \emph{Applicable Scenarios}, \emph{Characteristics}, and \emph{Good Examples}, where \emph{Good Examples} include representative stego text with the corresponding metric scores. Across iterations, we observe a clear progression: early strategies yield limited embedding capacity, lower text quality, and weaker anti-steganalysis performance, whereas later strategies achieve higher embedding rate, better fluency and coherence, and improved  statistical invisibility. The strategy evolution process in Table \ref{tab:strategy_evolution} shows that Auto-Stega can autonomously develop effective strategies for text steganography, underscoring its practical value for covert communication.

\label{appendix2}
\begin{table*}[t]
  \centering

  \begingroup
  \small
  \setlength{\tabcolsep}{2.5pt}
  \renewcommand{\arraystretch}{1.45}

  \newcommand{\celll}[2]{\parbox[c]{#1}{\raggedright #2}} 
  \newcommand{\cellc}[2]{\parbox[c]{#1}{\centering #2}}   
  \newcommand{\thickhline}{\noalign{\hrule height 1.2pt}}
  \newcommand{\midline}{\noalign{\hrule height 0.6pt}}

  \begin{minipage}{0.90\linewidth}
    \centering
    \begin{tabular}{@{}l l l l l l c@{}}
      \thickhline
      \cellc{0.16\linewidth}{\textbf{Strategy}} &
      \cellc{0.20\linewidth}{\textbf{Core idea}\\\textbf{(one sentence)}} &
      \cellc{0.09\linewidth}{\textbf{Technique}} &
      \cellc{0.09\linewidth}{\textbf{Scenarios}} &
      \cellc{0.12\linewidth}{\textbf{Key traits}} &
      \cellc{0.18\linewidth}{\textbf{Representative}\\\textbf{stego excerpt}} &
      \cellc{0.05\linewidth}{\textbf{Score}} \\
      \midline

      \celll{0.16\linewidth}{Thematic Domain Diversification with\\ Consistent Transitional Enc} &
      \celll{0.20\linewidth}{Combine semantic preservation with light rewriting to smooth statistical signals and reduce detectability while balancing capacity and robustness (esp.\ back-translation).} &
      \celll{0.09\linewidth}{semantic} &
      \celll{0.09\linewidth}{robustness; low\\ detectability (LLM)} &
      \celll{0.12\linewidth}{semantic preservation; back-translation robust; hard for detectors} &
      \celll{0.18\linewidth}{\emph{``Quantum computing will usher in a new era \ldots{} optimize trading strategies by processing market data \ldots{}''}} &
      \cellc{0.05\linewidth}{6.125} \\
      \midline

      \celll{0.16\linewidth}{Domain-Specific Semantic Precision\\ with Transitional Depth} &
      \celll{0.20\linewidth}{Use domain-specific, non-generic elaborations with consistent transitional phrases; varied transitions align with encoded bits and thematic depth to raise naturalness and lower detectability.} &
      \celll{0.09\linewidth}{Context-aware\\ semantic embedding} &
      \celll{0.09\linewidth}{general} &
      \celll{0.12\linewidth}{data-driven; non-generic;\\ varied transitions} &
      \celll{0.18\linewidth}{\emph{``Moreover, the implementation of quantum-resistant cryptographic algorithms necessitates evaluation of lattice-based schemes \ldots{}''}} &
      \cellc{0.05\linewidth}{7.625} \\
      \midline

      \celll{0.16\linewidth}{Domain-Specific Semantic Depth\\ with Multi-Domain Redundancy} &
      \celll{0.20\linewidth}{Embed via consistent transitions plus rich, domain-adapted elaborations across multiple domains; introduce paragraph-level redundancy to improve error tolerance without generic phrasing.} &
      \celll{0.09\linewidth}{context-aware\\ semantic embedding} &
      \celll{0.09\linewidth}{general} &
      \celll{0.12\linewidth}{data-driven; semantic depth;\\ redundancy; low detectability} &
      \celll{0.18\linewidth}{\emph{``Moreover, in smart grid management, advanced metering enables real-time demand response \ldots{} enhancing grid stability \ldots{}''}} &
      \cellc{0.05\linewidth}{9.000} \\
      \thickhline
    \end{tabular}
  \end{minipage}

  \endgroup
\caption{Strategy evolution summary for stego text generation (concise view).}
\label{tab:strategy_evolution}
\end{table*}

\subsection{Full prompts for the Steganography LLM and Summarizer LLM}
\label{appendix3}
In this section, we present the prompt templates that operationalize the two core components of Auto-Stega. The Summarizer LLM is prompted to distill observed improvements into structured strategy entries under strict output and safety constraints, as shown in Fig. \ref{prompt}. During the warm up stage, the steganographic LLM embeds a given secret into a cover subject to fluency and coherence requirements, with controlled length variation and avoidance of statistical artifacts, as shown in Fig. \ref{prompt_steganography1}. During the lifelong learning stage, the generator conditions on retrieved strategy headers, either a single strategy or a composed set, and applies them to produce candidates under the same quality requirements, as shown in Fig. \ref{prompt_steganography2}. These templates standardize inputs and outputs across iterations, enable compositional use of strategies, and support reliable extraction and consistent evaluation throughout the Auto-Stega framework.

\definecolor{codebg}{RGB}{235,248,238}
\definecolor{numgray}{gray}{0.45}
\definecolor{rulegray}{gray}{0.75}
\lstdefinestyle{pygreen}{
  language=Python,
  backgroundcolor=\color{codebg},
  basicstyle=\ttfamily\small,
  numbers=left,
  numberstyle=\footnotesize\color{numgray},
  stepnumber=1,
  numbersep=10pt,
  frame=single,
  framerule=0.6pt,
  rulecolor=\color{rulegray},
  showstringspaces=false,
  keepspaces=true,
  breaklines=true,
  columns=fullflexible
}

\begin{figure*}[t]
  \centering
  \begin{minipage}{0.96\linewidth}
\begin{lstlisting}[style=pygreen]
def strategy_summarizer_prompt():
    return (
"You are a steganography strategy summarizer.\n"
"Think through the task step by step INTERNALLY and summarize your reasoning, but DO NOT reveal your reasoning.\n"
"Output ONLY a valid JSON object and nothing else.\n"
"\n"
"When unsure or if required information is missing, return the empty JSON object: {}.\n"
"\n"
"=== Output contract ===\n"
"Return a JSON object with EXACTLY these keys in this order:\n"
"  name (string),\n"
"  definition (string),\n"
"  technique (array of strings),\n"
"  applicable_scenarios (array of strings),\n"
"  characteristics (array of strings),\n"
"  examples (array of objects with keys: stego_excerpt (string), overall_score (number), scores (object)).\n"
"- Use double quotes for all strings; no trailing commas; UTF-8 text.\n"
"- Match the output language to the cover_text language if detectable; otherwise use English.\n"
"- Never include secrets or secret_bits_preview in the output; never echo raw secret bits.\n"
"\n"
"=== Internal step-by-step plan (do not expose) ===\n"
"1) Parse inputs (stego_text, cover_text, evaluation, used_strategies). Identify signals such as: "
"equiprobable LM, acrostic AM, cover continuation, bin size, temperature, decoding hints, etc.\n"
"2) Draft a concise, distinctive NAME (<= 60 chars, Title Case). Prefer pattern: "
"'<Core Mechanism> + <Key Modifier> (optional params)'.\n"
"3) Write a 1-2 sentence DEFINITION that states mechanism (how), purpose (why), and limits (if any).\n"
"4) TECHNIQUE: list concrete methods/ingredients (e.g., 'equiprobable token binning (bin=64)', "
"'deterministic acrostic A-M->0/N-Z->1', 'cover-conditioned continuation', 'low-temp decoding').\n"
"5) APPLICABLE_SCENARIOS: list contexts where this strategy is appropriate (e.g., 'short social posts', "
"'news-style paragraphs', 'low-detectability requirement', 'receiver knows cover').\n"
"6) CHARACTERISTICS: list crisp traits, mixing strengths/risks/requirements (e.g., 'capacity approx 6 bpw', "
"'good fluency under low temperature', 'robustness limited by rephrasing', 'needs shared cover').\n"
"7) EXAMPLES: choose 1-3 short stego excerpts from stego_text (<= 240 chars each, no ellipsis if avoidable). "
"For each example, set overall_score and scores using evaluation if available; otherwise make a conservative estimate "
"from provided fields. Round numbers to one decimal and clamp to [0,10].\n"
"\n"
"=== Scoring normalization (internal) ===\n"
"- If evaluation provides per-dimension scores in [0,10], copy directly.\n"
"- If given probabilities (e.g., detection probability in [0,1]), map to score as (10 - 10*p), rounded to one decimal.\n"
"- If only raw metrics exist (e.g., PPL), use any provided *_score fields; do NOT invent unseen dimensions.\n"
"- If no usable signals exist for scoring, output the empty JSON object {}.\n"
"\n"
"=== Safety & formatting rules ===\n"
"- Do not include explanations, headings, or Markdown -- JSON only.\n"
"- Do not fabricate technical parameters that are not implied by inputs.\n"
"- Express the strategy unambiguously; given the stego_text and its corresponding strategy description, the embedded secret bits can be decoded with 100% reliability.\n"
)
\end{lstlisting}
  \end{minipage}
\caption{Prompts used by the Summarizer LLM in Auto-Stega.}
\label{prompt}
\end{figure*}

\begin{figure*}[t]
  \centering
  \begin{minipage}{0.96\linewidth}
\begin{lstlisting}[style=pygreen]
def base_generate_prompt(secret, cover_stub, method_line=""):
    return f"""
System:
You are a professional text steganography expert. Your task is to embed a secret naturally into carrier text.

User:
You are participating in a steganography research project. Your goal is to produce high-quality stego text. Embed the given secret while keeping the text natural and fluent.

Requirements:
1) Secret bitstring to embed: {secret}
2) Carrier text: {cover_stub}
3) Keep the text natural and fluent; avoid statistical artifacts
4) Limit length change to within +/- 10% (if you write the carrier yourself, target 80-200 words)
5) You may use covert techniques such as synonym substitution, syntax restructuring, and context-aware semantic embedding{method_line}

Core principles:
- Maintain semantic consistency and topical coherence
- Avoid suspicious anomalies (e.g., odd token/character distributions)
- Ensure the secret can be correctly extracted

Produce the stego text only, and wrap it between [START STEGO TEXT] and [END STEGO TEXT].
"""
\end{lstlisting}
  \end{minipage}
\caption{Prompts used by the Steganography LLM during the warm-up stage of Auto-Stega.}
\label{prompt_steganography1}
\end{figure*}

\begin{figure*}[t]
  \centering
  \begin{minipage}{0.96\linewidth}
\begin{lstlisting}[style=pygreen]
def strategy_header_single(name, definition, technique, example):
    return (
        "Please use the following steganography strategy:\n"
        f"Strategy Name: {name}\n"
        f"Definition: {definition}\n"
        f"Technique: {technique}\n"
        f"Example: {example}"
    )

def strategy_header_multi(items):
    # items: list of dicts with keys: Strategy, Technique, Definition
    lines = []
    for s in items:
        lines.append(
            f"- {s.get('Strategy','unknown')} ({s.get('Technique','unknown')})\n"
            f"  Definition: {s.get('Definition','')}"
        )
    return "Please combine the following strategies:\n" + "\n".join(lines)

def with_strategy_prompt(secret, cover_stub, strategies_desc, method_line=""):
    return f"""
System:
You are a professional text steganography expert. Apply the specified strategies to embed the secret.

User:
You are participating in a steganography research project. Apply the specified strategy/strategies to produce stego text.

{strategies_desc}

Requirements:
1) Secret bitstring to embed: {secret}
2) Carrier text: {cover_stub}
3) Keep the text natural, coherent, and statistically unremarkable
4) Strictly apply the specified strategy/strategies for embedding{method_line}

Produce the stego text only, and wrap it between [START STEGO TEXT] and [END STEGO TEXT].
"""
\end{lstlisting}
  \end{minipage}
\caption{Prompts used by the Steganography LLM during the lifelong learning stage of Auto-Stega.}
\label{prompt_steganography2}
\end{figure*}

\end{document}